\documentclass[12pt,a4paper,notitlepage]{article}
\usepackage[utf8x]{inputenc}
\usepackage{ucs}
\usepackage{amsmath}
\usepackage{amsfonts}
\usepackage{amssymb}
\usepackage{graphicx}

\usepackage{authblk}

\DeclareUnicodeCharacter{8208}{}

\title{First-principles study of electronic and optical properties in 1-D oligomeric derivatives of telomestatin}

\author[1]{Jo\"{e}lle M\'{e}rgola-Greef}
\author[1,2]{Bruce F. Milne\thanks{bruce.milne$@$abdn.ac.uk}}
\affil[1]{Marine Biodiscovery Centre, Department of Chemistry, University of Aberdeen, Meston Building, Meston Walk, AB24 3UE, Old Aberdeen, UK.}
\affil[2]{CFisUC, Department of Physics, University of Coimbra, Rua Larga, 3004-516 Coimbra, Portugal.}

\begin{document}
\maketitle

\begin{abstract}
Real-space self-interaction corrected (time-dependent) density functional theory has been used to investigate the ground-state electronic structure and optical absorption profiles of a series of linear oligomers inspired by the natural product telomestatin. Length-dependent development of plasmonic excitations in the UV region is seen in the neutral species which is augmented by polaron-type absorption in the IR when the chains are doped with an additional electron/hole. Combined with a lack of absorption in the visible region this suggests these oligomers as good candidates for applications such as transparent antennae in dye-sensitised solar energy collection materials.
\end{abstract}

\section{Introduction}

Organic materials (OM) have seen huge increases in optical and electronic applications in recent years.\cite{Hu1999,Shi2021,Zhao2013} The list includes flexible/wearable electronic devices, organic light-emitting diodes (OLED), organic field-effect transistors (OFET), (bio)sensors and the harvesting, storage and transmission of energy.\cite{Sun2017,Machin2021,Saad2021,Minder2012} In addition, there is great interest in OM for electron spin-manipulation in quantum data processing and storage.\cite{Majumdar2006,Bubnova2014}

In particular, 1-dimensional (1D) components approaching the quantum confinement limit of atomic chains have received considerable attention due to their unique properties.\cite{Bryant2016,Yan2007} Atomic chains themselves are the ideal model for this but in practice are too limited for real-world use. Pseudo-1D systems such as carbon nanotubes show considerable promise but again are limited in applicability due to difficulties in controlling their geometrical and electronic structure in order to realise predicted properties.\cite{Silva2020,Silva2018,Bao2018}

Organic polymer/oligomer systems based on $\pi$-conjugated heterocycles with applications in (opto)electronics are well known (e.g. polypyrroles, polythiophenes, polyacenes).\cite{Majumdar2006,Fernandes2018,Lauchner2015} These closely approach the 1D limit and great effort has gone into tailoring their properties for usable device applications. Unfortunately, the modifications inevitably lead to considerable alterations in the size of the basic monomer units which in turn change the charge/spin carrier density profile and also alter the machinability of the materials.\cite{Bao2018,Kimpel2022}

\begin{figure}[!h]
    \centering
    \includegraphics[scale=1]{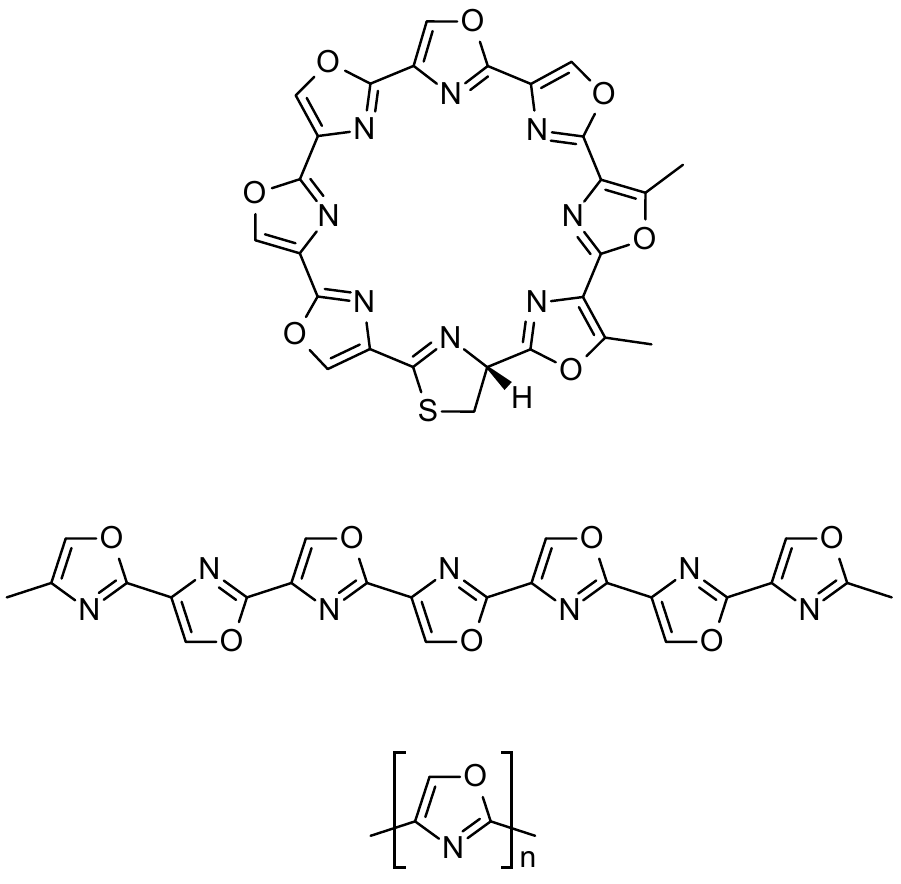}
    \caption{The all-\textit{cis} natural product \textit{R}-telomestatin (top), linear all-\textit{trans} model based on telomestatin (middle) and general schematic of the methyl-capped linear 1,4-polyoxazole oligomers studied in this work (bottom).}
    \label{fgr:telomestatin}
\end{figure}

Natural products (NP) have long been a source of inspiration for chemists, particularly in the drug-discovery field. In materials science, NP tend to have their greatest influence in areas such as biomaterials based on soft matter and macromolecular biomimetics.\cite{Kohnke1989,Sanchez2005,Ige2012} 

The NP \textit{R}-telomestatin (Figure \ref{fgr:telomestatin}) is a macrocyclic compound produced by the organism \textit{Streptomyces anulatus} 3533-SV4 which has shown to display specific binding to the telomeric G-quadruplex DNA structure and and has strong telomerase inhibitory activity.\cite{Shinya2001,Kim2002}  Telomestatin is composed of seven conjugated oxazole rings with macrocycle closure by a single thiazoline heterocycle. This structure is a ribosomally-synthesised, post-translationally modified peptide (RiPP) with the original precursor sequence being CTTSSSSS. In the biosynthetic pathway, this sequence has been found to be incorporated into a longer precursor peptide, TisC, which is then processed by various enzymes to yield the final product.\cite{Amagai2017}

The similarity between the basic structural features of telomestatin and known organic optoelectronic materials such as polypyrrole and polythiophene is striking. Furthermore, recent developments in the enzymatic modification of synthetic peptides to produce azole rings such as oxazole, thiazole and selenothiazole in the laboratory suggest that polymeric structures based on these motifs can be produced in a high-yield, environmentally friendly way.\cite{Koehnke2013a,Koehnke2013b,Houssen2014}

The current work starts with a summary of the theoretical methods employed followed by an investigation of the ground- and excited-state electronic properties of a series of linear oligomeric oxazole chains inspired by the telomestatin structure. We begin with an analysis of the static electric polarisability of the oligomers where, using highly-converged MP2 calculations, we benchmark our self-interaction corrected TDDFT (SIC-TDDFT) method. We then further investigate the ground-state properties of the polyoxazoles through the evolution of their electronic density of states (DOS) to gain a insight into changes in the electronic structure with chain length. Finally, the optical absorption properties of the chains are studied using time-propagation SIC-TDDFT to probe the excited-state electronic behaviour of the oligomers in their neutral and electron-/hole-doped states.


\section{Theory}
\subsection{Real-space DFT}
Density functional theory (DFT),\cite{Hohenberg1964,Kohn1965} using (semi-)local functionals of the electronic charge density is well suited to discretization on real-space grids.\cite{Nogueira2003} In the current work the \textsc{Octopus} code is used which employs regular spatial grids within a user-defined volume of space for this purpose and does not make use of a basis-set expansion of the Kohn-Sham wavefunctions with atom-centered or plane-wave basis functions.\cite{Octopus2020,Octopus2015,Octopus2014,Octopus2012} 

Convergence of calculated properties thus depend only on the grid spacing and the size of simulation box chosen. This has the added convenience that real-space calculations of this type are inherently scalable on (massively) parallel computing platforms and therefore able to provide the same level of accuracy for extremely large systems as might be expected for small molecules using quantum chemical DFT implementations.\cite{Jornet-Somoza2015}

In the following, electron spin is neglected in order to simplify notation but these equations can easily be generalized to include this additional degree of freedom.

We solve the ground-state (time-independent) Kohn-Sham system of $N$ coupled one-particle Schr\"odinger equations

\begin{equation}
    h_{\mathrm{KS}}\varphi_m(\textbf{r}) = \epsilon_m\varphi_m(\textbf{r})~~~~~(m = 1,\ldots,N)
    \label{gsks}
\end{equation}

\begin{equation}
    h_{\mathrm{KS}}[n] = -\frac{1}{2}\nabla^2 + v_{\mathrm{KS}}(\textbf{r})
\end{equation}

\noindent which in turn allows us to compute the total charge density $n(\textbf{r})$ and total energy $E_{\mathrm{total}}[n(\textbf{r})]$ for our system

\begin{equation}
    n(\textbf{r}) = \sum_m^N |\varphi_m(\textbf{r})|^2
\end{equation}

\begin{equation}
    E_{\mathrm{total}}[n(\textbf{r})] = \sum_m^N \epsilon_m
\end{equation}

The Kohn-Sham potential $v_{\mathrm{KS}}$ is comprised of the following components

\begin{equation}
    v_{\mathrm{KS}} = v_{\mathrm{ext}}(\textbf{r}) + v_{\mathrm{H}}[n(\textbf{r})] + v_{\mathrm{xc}}[n(\textbf{r})]
    \label{vks}
\end{equation}

For a molecule in the absence of applied electrical or magnetic fields the external potential $v_{\mathrm{ext}}(\textbf{r})$ is simply the net Coulomb potential due to the positively charged nuclei. In real-space DFT it is convenient to separate the chemically distinct valence and core charge densities and make use of pseudopotentials to describe the core electron density since this region of space would otherwise require very dense spatial grids to accurately describe the rapid density fluctuations found near the nucleus.\cite{Nogueira2003,Austin1962,Schwerdtfeger2011} Thus, we reduce the full electronic structure problem to one with the valence electrons moving in an external potential due to the positively charged core ions.

The classical electrostatic mean-field interaction between the electrons is described by the Hartree potential $v_{\mathrm{H}}[n(\textbf{r})]$ which is conventionally written as

\begin{equation}
    v_{\mathrm{H}}[n(\textbf{r})] = \int \mathrm{d^3}\textbf{r}'\frac{n(\textbf{r}')}{|\textbf{r} - \textbf{r}'|}
    \label{vH_int}
\end{equation}

Whilst it is convenient to evaluate this term as an integral when using e.g. atom-centered Gaussian basis functions for small systems, with real-space grids and large spatial dimensions this rapidly becomes computationally challenging. In the \textsc{Octopus} calculations reported here $v_{\mathrm{H}}[n(\textbf{r})] $ is obtained by solving Poisson's equation

\begin{equation}
    \nabla^2 v_{\mathrm{H}}[n(\textbf{r})] = -4\pi n(\textbf{r})
    \label{poisson}
\end{equation}

The final term in Eqn \ref{vks} is the exchange-correlation potential $v_{xc}[n(\textbf{r})]$

\begin{equation}
    v_{xc}[n(\textbf{r})] = \frac{\partial E_{xc}[n(\textbf{r})]}{\partial n(\textbf{r})}
\end{equation}

 which accounts for all remaining quantum mechanical and Coulomb repulsion effects. As the exact form of the universal exchange-correlation functional $E_{xc}[n(\textbf{r})]$ is not known, this must be approximated.


\subsection{Real-time TDDFT}
Once the ground state orbitals $\varphi_m(\textbf{r})$ have been obtained using the real-space DFT scheme outlined above we can insert them into the time-dependent Kohn-Sham equation
\begin{equation}
    i\frac{\partial }{\partial t}\varphi_m(\textbf{r}, t) = \left[ -\frac{1}{2}\nabla^2 + v_{\mathrm{KS}} [n(\textbf{r}, t)] \right] \varphi_m(\textbf{r}, t)
    \label{tdks}
\end{equation}

\noindent which yields the time-dependent density $n(\textbf{r}, t)$ of the system

\begin{equation}
    n(\textbf{r}, t) = \sum_m^N |\varphi_m(\textbf{r}, t)|^2
\end{equation}

We can now allow this to evolve in real time to obtain information on time-dependent properties such as electronic excitations. To do this we propagate the electronic states $\varphi_m$ and storing the resulting total density at a given point in time gives us access to the time-dependent density $n(\textbf{r}, t)$. However, in order to extract time-dependent information we need to excite the system in some way (otherwise no change would be seen and we would have $n(\textbf{r}, t) = n(\textbf{r})$ at all points in the propagation).

If we are working with small perturbations such as those that would be found in ordinary UV-Vis absorption processes, we can assume that the processes are linear with respect to the size of the perturbing force and apply a small impulse (or 'kick') to add a finite momentum $\mathcal{K}$ to the electrons

\begin{equation}
\varphi_m (\textbf{r}, \delta t) = e^{i\mathcal{K}z}\varphi_m (\textbf{r}, 0)
\end{equation}

\noindent which excites all modes of the electronic system equally (in this example we are propagating in the $z$-direction). This is applied at t = 0 and immediately switched off for the rest of the propagation so that the response of the density to this perturbation can be monitored. 

In the case where we are interested in optical absorption, we can use the dipole approximation and save the induced dipole $\mu(t)$ at various points throughout the propagation. This dipole trajectory can then be Fourier transformed to obtain the frequency-dependent dynamic polarisability, $\alpha(\omega)$, of the system

\begin{equation}
\alpha(\omega) = \frac{1}{\mathcal{K}} \int dt ~e^{i \omega t} \left[\mu (t) - \mu (0) \right]
\label{eq:TDalpha}
\end{equation}

\noindent $\alpha(\omega)$ consists of both real and imaginary parts corresponding to scattering and absorption processes, respectively. We can extract the absorption spectrum by calculating the strength function, S$(\omega)$, which provides information on the magnitude of the response at a given frequency

\begin{equation}
\text{S}(\omega) = \frac{2\omega}{\pi} \operatorname{Im} \alpha(\omega)
\end{equation}

\noindent The plot of S($\omega$) vs $\omega$ corresponds to the absorption spectrum of the system.

\subsection{Sternheimer linear response}
Solution of the TDDFT equations in the time domain provides an efficient route to the full dipole response spectrum, as shown above. However, whilst the accuracy of the polarisabilities obtained is good enough for prediction of optical absorption spectra it is dependent on the length of the propagations used. More accurate polarisability values for a given frequency can be obtained from perturbation theory with less computational effort but the cost rises rapidly with the number of frequencies required, limiting its practical use to small numbers of discrete frequencies.

Within the framework of TDDFT, the Sternheimer equation serves this purpose and permits the calculation of frequency-(in)dependent (hyper)polarisabilites using only the occupied Kohn-Sham orbitals, $\varphi_m$ and without the need for infinite sums over unoccupied states.\cite{Sternheimer1954,Mahan1980} For a monochromatic perturbing field $F = \lambda\delta v(\textbf{r}) cos(\omega t)$ we get (to first-order) the variation of the density as

\begin{equation}
\delta n(\textbf{r}, \omega) = \sum_m^{N} \bigg\{ [\varphi_m(\textbf{r})]^* \delta\varphi_m(\textbf{r}, \omega) + [\delta\varphi_m(\textbf{r}, -\omega)]^* \varphi_m(\textbf{r}) \bigg\}
\end{equation}

The perturbed Kohn-Sham orbitals $\delta\varphi_m$ are used to construct the Sternheimer perturbation expression

\begin{equation}
\lbrace \hat{h}_{KS} - \epsilon_m \pm \omega + i\eta \rbrace \delta\varphi_m(\textbf{r}, \pm\omega) = -P_c \delta \hat{h}_{KS}(\pm \omega) \varphi_m(\textbf{r)}
\label{eq:sternheimer}
\end{equation}

\noindent where $P_c$ is a projector which removes $\delta \varphi_m (\textbf{r}, \pm\omega)$ components from the occupied subspace and $\eta$ is a positive infinitesimal that ensures convergence at or near resonance frequencies.\cite{Andrade2007}

Once $\delta n(\textbf{r}, \omega)$ has been obtained from self-consistent solution of Eq. \ref{eq:sternheimer}, the polarisability tensor $\alpha_{ij}$ can be calculated as

\begin{equation}
\alpha_{ij} = \int d^3 \textbf{r} ~\delta n_j(\textbf{r}, \omega) F_i
\end{equation}




\section{Methods}
\subsection{Geometry optimization}
Geometries were optimized at the DFT level using the Orca software package (version 5.0.3).\cite{Neese2012} The re-regularized SCAN meta-GGA functional (r$^2$SCAN) was chosen as it has been shown to provide equal or better geometries for diverse molecular systems when compared with more computationally expensive hybrid functionals and is competetive with correlated wavefunction methods.\cite{Sun2015,Furness2020} Although r$^2$SCAN does include short- to  intermediate-range VdW effects, as with all other (semi-)local functionals it is not capable of describing long range dispersion interactions. For this reason the accurate and computationally efficient density-dependent D4 dispersion correction was added to r$^2$SCAN during all geometry calculations.\cite{Ehlert2021,Caldeweyher2019} 

The r$^2$SCAN-D4 method was combined with the Def2-TZVP basis set.\cite{Schafer1992, Weigend2005} Although the original tests\cite{Ehlert2021} of the accuracy of r$^2$SCAN-D4 with respect to geometrical parameters were performed with the more complete Def2-QZVP basis, in this work it was found that the Def2-TZVP basis lead to mean absolute deviations relative to the Def2-QZVP basis of approximately 0.001 \AA ~in bond lengths (maximum 0.002 \AA) and 0.002 degrees in bond angles (maximum 0.060 degrees) for a test system of the methyl-capped oxazole dimer. It was therefore concluded that the r$^2$SCAN-D4/Def2-TZVP method would be sufficiently accurate for this study.

\subsection{Time-propagation TDDFT}
As stated in the Theory section we employed real-space grids for the discretization of the (TD)DFT equations with the \textsc{Octopus} code version 12.0. The exchange and correlation terms were approximated using the modified Perdew-Zunger local-density approximation (LDA).\cite{Dirac1930,Bloch1929,Perdew1981} This was provided by the LibXC library version 5.2.3.\cite{Marques2012,Lehtola2018}

The LDA calculations were augmented by including the average-density self-interaction correction (ADSIC) which corrects the Kohn-Sham potential $v_{KS}$ to remove the spurious interactions of each electron with itself which arise from the Hartree term (see Eqn \ref{vH_int}). The correction to the potential for a system containing $N$ electrons is obtained as

\begin{equation}
    v_{ADSIC} = v_{KS} - \left[ v_H \left( \frac{n(\textbf{r})}{N} \right) + v_{xc} \left( \frac{n(\textbf{r})}{N} \right) \right]
\end{equation}

\noindent and restores the correct $-1/r$ asymptotic decay of the charge density.\cite{Legrand2002}

The addition of the self-interaction correction is particularly important for the current work as we were interested in the effect of doping of the compounds under study with added electrons and the uncorrected LDA fails to yield bound anionic systems. ADSIC corrects this failure and in all cases studied here the additional electron was bound indicating that the anions were stable.

Scalar relativistic pseudopotentials for the ADSIC-LDA calculations were taken from the standard norm-conserving LDA set available from the Pseudo Dojo website and distributed with \textsc{Octopus}.\cite{vanSetten2018,Hamann2013,Hamann2017,Garrity2014} Spin-orbit contributions were not considered as these were expected to be insignificant in the context of the current study. 

Convergence testing of the grid spacing and the radius of the atom-centred spheres used to construct the simulation box was performed for the linear six-membered oxazole oligomer. It was found that with a spacing of 0.2 \AA ~and radius of 6 \AA, the Fermi energy (E$_{\mathrm{Fermi}}$) of the neutral species was converged to better than 0.01 eV. Upon varying the electron number, for monocationic species the same combination of spacing and radius was used but in order to obtain the same convergence of E$_{\mathrm{Fermi}}$ it was necessary to double the radius to 12 \AA.

All time-propagation TDDFT calculations of absorption spectra used a time step of 0.003 $\hbar ~\mathrm{eV}^{-1}$ (0.00197 fs) which was found to yield stable propagations and maintain energy conservation. The propagations were carried out for 8000 steps to give a total time of 24 $\hbar ~\mathrm{eV}^{-1}$ (15.8 fs)

\subsection{Static polarisability calculations}

Calculation of static polarisabilities, $\alpha_0$, for the various oligomer chains was performed at the TDDFT level (LDA and ADSIC-LDA) by solving the Sternheimer equation for $\omega = 0$.\cite{Andrade2007} As the chains studied were neutral, the grid spacing of 0.2 \AA ~and radius of 6 \AA ~were used. In order to ensure accurate results in the perturbation calculations, the electronic states were converged more tightly by setting the eigensolver tolerance to $10^{-10}$ Hartree.

For comparison, benchmark values for the polarisabilities were obtained using second-order M\o{}ller-Plesset (MP2) perturbation theory with the Orca software. In order to be consistent with the TDDFT polarisability caluclations, scalar relativistic effects were included in the MP2 calculations using the zero-order regular approximation (ZORA).\cite{Faas1995, VanLenthe1998} Convergence of the polarisabilities was tested with the sequence of polarised basis sets ZORA-Def2-SVP, ZORA-Def2-TZVP and ZORA-Def2-QZVPP. As the diffuse-augmented ZORA basis sets were not available internally in the Orca software, the uncontracted diffuse functions from the corresponding Def2-SVPD, Def2-TZVPD and Def2-QZVPPD basis sets were obtained from the Basis Set Exchange (https://www.basissetexchange.org/) and added to the recontracted ZORA basis sets.\cite{BSC2019}. The SARC auxiliary basis sets were used along with the resolution of the identity (RI) approximation for the Hartree-Fock calculations.\cite{Weigend2006} Similarly, the subsequent MP2 correlation treatment also used the RI approximation and employed the internal Def2/C auxiliary basis sets.\cite{Hellweg2007}

\section{Results and discussion}
\subsection{Static polarisabilities}

\begin{figure}
\centering
\includegraphics[scale=1]{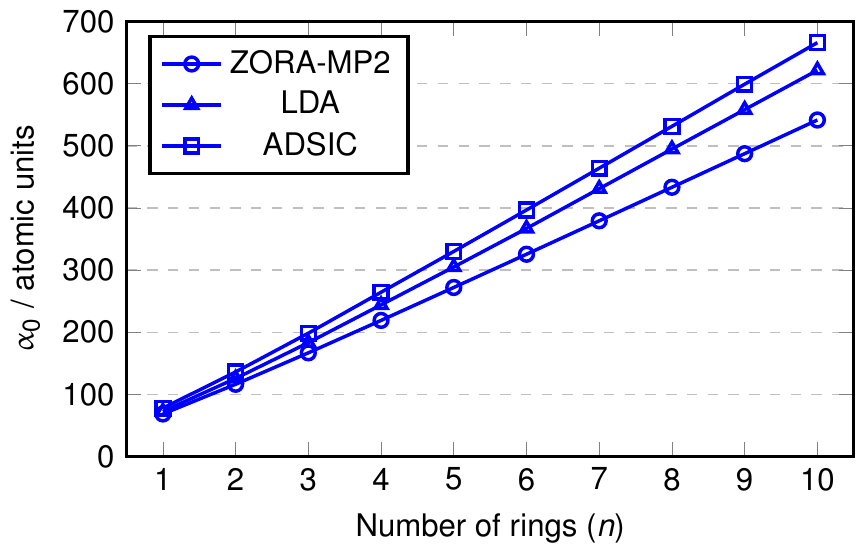}

\caption{Static isotropic polarizability ($\alpha_0$) of oxazole chains. Lines added between points to emphasize trends.}
\label{alpha}
\end{figure}

\begin{figure}[h!]
\centering
\includegraphics[scale=1]{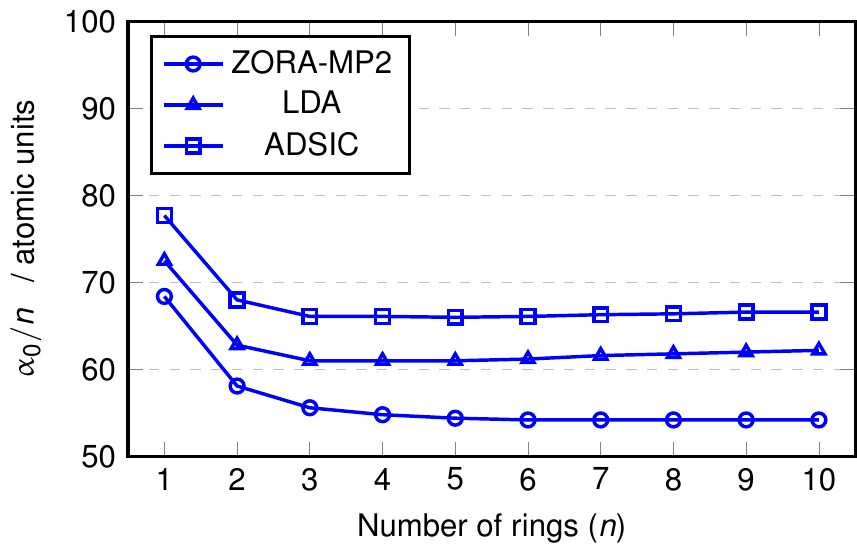}

\caption{Static isotropic polarizability ($\alpha_0$) of oxazole chains divided by number of rings ($n$). Lines added between points to emphasize trends.}
\label{alpha_n}
\end{figure}

TDDFT using conventional (semi)local density functional approximations such as LDA or GGA is known to overestimate the polarisabilities of conjugated molecular chains which could adversely impact the calculation of their optical properties (see Equation \ref{eq:TDalpha}).\cite{Marques2003,deBoeij2006,Champagne1998,VanGisbergen1999} For this reason we performed calculations of the static polarisability, $\alpha_0$, with the ADSIC method and compared them against benchmark MP2 values. As a further test, uncorrected LDA calculations were included in order to evaluate the difference made by the self interaction correction. All polarisability values were converged to $\sim$ 1 a.u (the MP2 values to $<$ 0.5 a.u.).

Figure \ref{alpha} shows the values of $\alpha_0$ obtained for the polyoxazole chains (n = 1--10). In all cases, the polarisability appears to increase in a linear fashion up to the chain with n = 10. The only deviation from this linearity appeared as a small change in slope at around n= 2 or 3, so the data was replotted as $\alpha_0$ divided by the number of rings in the chain (Figure \ref{alpha_n}). 

In this replotted form, it is clear that there is a large decrease in $\alpha_0 / n$ on formation of the initial inter-ring conjugation at n = 2, with smaller decreases seen as a consequence of adding subsequent rings to the core of the chain. The benchmark MP2 results indicate a clear convergence towards a fixed value of $\alpha_0 / n$ at approximately n= 5 and as expected, the uncorrected LDA data shows a gradual increase in $\alpha_0 / n$ as the chain length becomes larger. Conversely, the ADSIC data indicates that the self-interaction correction removes this spurious effect and leads to converged $\alpha_0 / n$ values and suggests that the correct physical picture is obtained with this method.

It is interesting to note that rather than improving on the absolute value of the LDA polarisability, ADSIC actually leads to slightly larger $\alpha_0 / n$ values. This is in contrast to the effect seen for SIC-LDA in simple hydrogen chains, however, these do not display the extended $\pi$-conjugation found in the polyoxazole oligomers.\cite{Pemmaraju2008} The overestimation relative to LDA seen here was small and judged to be of little consequence in relation to the present study, particularly because whereas LDA yielded unbound anionic species (i.e. the energy of added electrons was positive), ADSIC correctly gave bound anions for all chain lengths studied in the optical absorption calculations (see below) permitting study of both hole- and electron-doped species in addition to the neutral chains.

The benchmark MP2 calculations and the fact that the SIC-TDDFT results agree with them provides reliable evidence that the polyoxazole oligomers should display polarisabilities that may be modulated in a linear fashion simply by tuning the length parameter, n. This is expected to have important consequences for their electrical and optical properties and suggests that these are interesting candidates for organic materials applications.\cite{Gillet2021} 

In addition, the fact that the ADSIC calculations appear to give the correct physics with respect to this property (albeit with a small positive shift in $\alpha_0 / n$) was seen as evidence that the calculated optical absorption properties of the polyoxazole chains obtained at the ADSIC level would be similarly reliable.

\subsection{Density of states}

\begin{figure*}
\centering
\includegraphics[width=\textwidth]{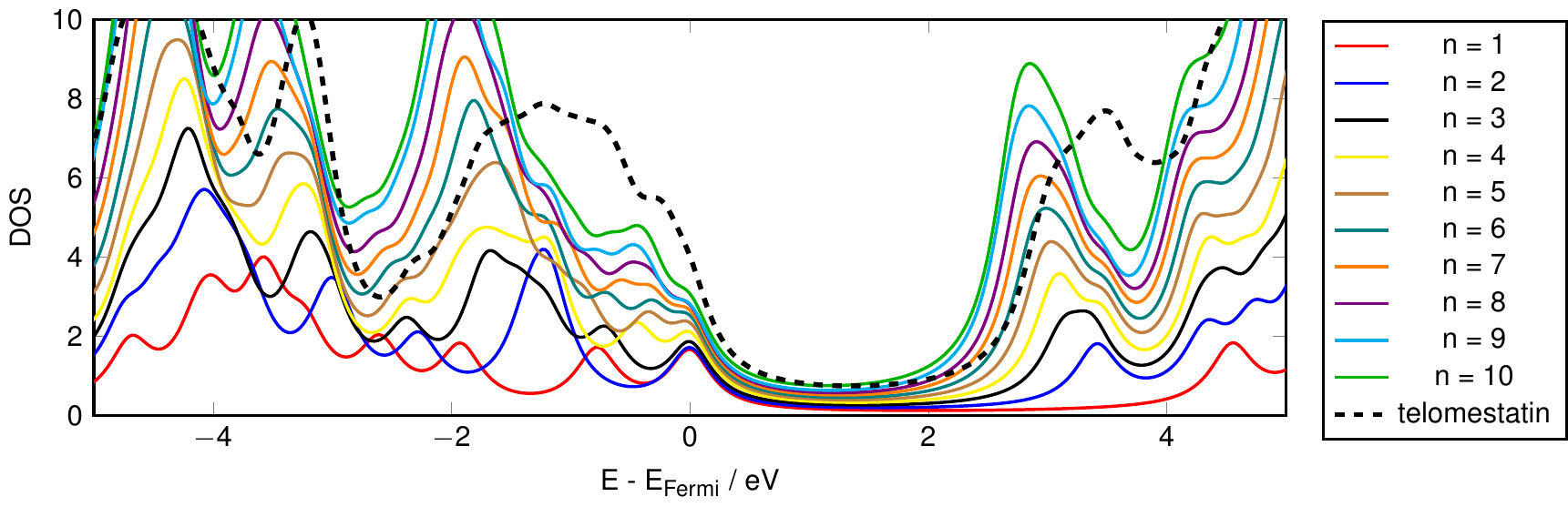}
\caption{TD-ADSIC density of state (DOS) plots for oligomeric oxazole chains. Energy scale normalised by shifting E$_{\mathrm{Fermi}}$ to zero for each oligomer. The non-zero value of the plot in the gap region is a result of the Lorentzian fit used to produce the plots but no states are present in this region.}
\label{fgr:dos}
\end{figure*}

The density of states calculated at the ADSIC level for the neutral oxazole oligomers are shown in Figure \ref{fgr:dos}. For clarity, the energy axis has been zeroed by subtracting the Fermi energy, E$_{\mathrm{Fermi}}$, from the state energies.

A large decrease in the gap energy can be seen ongoing from n = 1 to n=2, corresponding to the formation of the inter-ring conjugation in the dimer. Thereafter, a gradual red-shift in the peak energy accompanies the elongation of the oligomer chain along with a linear growth of the DOS corresponding to the development of a discrete conduction band (CB) centred at $\sim$ 3 eV above E$_{\mathrm{Fermi}}$. These data agree with the picture obtained from the changes in polarisability per ring ($\alpha_0 / n$) seen in the previous section.

Similarly, on going from n = 2 to n = 3, a split is seen in the CB peak. This is once again presumably due to the introduction of the third, central, oxazole ring and the continued existence of the n = 2 peak as a shoulder at $\sim$ 3.5 eV in the longer chains indicates that the terminal oxazoles contribute differently to the CB and that the main peak at $\sim$ 3 eV can be ascribed to the core heterocycles within the chain. This core/termini structure can be seen in the linear n = 7 telomestatin analogue in Figure \ref{fgr:telomestatin} 

The discrete nature of the CB peak in the longer oligomers indicates an increased density of CB states suggesting that these oligomers should display good electrical conductivity either when electrons are excited to the CB by optical absorption or by doping of the oligomers with excess electrons which will populate the lowest CB states.

The DOS obtained for telomestatin shows a different distribution of both the occupied valence region and the CB. These differences can be explained qualitatively by the all-\textit{cis} conformation of telomestatin versus the all-\textit{trans} conformation of the linear oligomers due to differences in strain on the conjugated $\pi$-system.


\subsection{Optical absorption spectra}
\subsubsection{Neutral oligomers}
\begin{figure*}
\centering
\includegraphics[width=\textwidth]{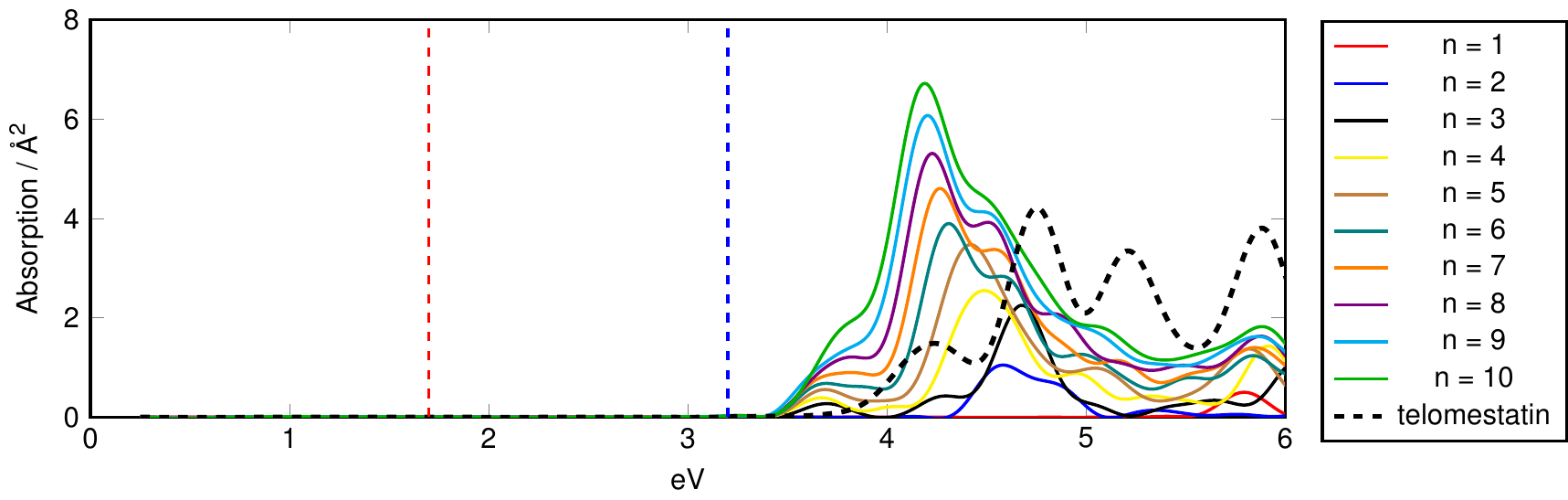}

\caption{Isotropic TD-ADSIC absorption spectra of neutral oligomeric oxazole chains. Vertical dashed lines indicate visible region. Spectra shifted by +0.25 eV.\cite{Doi2006}}
\label{fgr:abs_neutral}
\end{figure*}

The optical absorption spectra calculated at the TD-ADSIC level for the neutral oligomers are shown in Figure \ref{fgr:abs_neutral}. 


For comparison, the absorption spectrum of telomestatin was calculated at the same level of theory. When compared to the experimental UV-Vis spectrum of telomestatin (supporting information in Doi et al, 2006\cite{Doi2006}), the calculated spectrum was found to agree well in terms of the relative peak positions but showed a systematic underestimation of the peak energies. The solvent used in obtaining the experimental UV-Vis spectrum of telomestatin was not reported (given the other experimental details, however, it is likely that the final solvent in the synthesis, CH$_2$Cl$_2$ was used). It is reasonable to expect that the well-known underestimation of excitation energies by TD-LDA (even with the ADSIC correction) is likely to be significantly larger than minor alterations due to switching between nonpolar solvents capable of dissolving telomestatin, therefore justifying the use of the +0.25 eV shift in the context of the present work.

Adding a rigid shift of +0.25 eV to the calculated spectrum to match the energy of the dominant experimental peak gave an excellent fit to the experimental spectral features and resulted in errors of $\sim$ 0.1 eV in the remaining peak positions. Because of the good fit of the shifted TD-ADSIC data to the experimental spectrum of the natural product and the chemical similarity of the linear oligomers, the same rigid shift was subsequently applied to all other calculated spectra.

Looking at the spectra for the linear chains Figure \ref{fgr:abs_neutral}, it can be seen that the simplest system (n = 1) displays a single peak at 5.8 eV. Consistent with formation of the inter-ring $\pi$-conjugation for n = 2, the main peak shifts to 4.6 eV with formation of a shoulder approximately 0.3 eV above this.. At n = 3, these two features reverse relative intensity with the higher energy peak becoming largest. As was seen for the polarisability and DOS, above, on going to chains with n $>$ 3, the main peak is seen to gradually undergo a red-shift as the chain length increases. This shift can be seen to be converging towards a value of $\sim$ 4.1 eV suggesting that further growth of the conjugated core should not significantly impact the energy of this peak.


Unlike the peak energy, the peak intensity continues to grow in a linear fashion up to the maximum chain length studied here (n = 10). This is indicative of a plasmonic collective excitation which begins in the chains with n $>$ 2 (i.e. those with the core-termini structure) and suggests that both the energy and intensity of this plasmonic mode of the polyoxazole chains can be modulated by controlled synthesis of oligomers of a specific length.

Interestingly, the intensities of the first two absorption peaks for telomestatin are almost identical to those of the linear chain with the same number of oxazole rings (n = 7). A clear difference between the two compounds can be seen in the peak energies which are blue-shifted by approximately 0.5 eV. This is in line with increased strain on the conjugated $\pi$-system in the all-\textit{cis} macrocyclic telomestatin and suggests that the absorption spectra of the acyclic compounds should be sensitive indicators of conformational change. Combined with the intense absorption associated with plasmonic excitations, this may have applications in sensing of metal ions where binding to e.g. the polyoxazole nitrogens can be expected to induce alterations in the \textit{cis/trans} balance within the chain.

\subsubsection{Electron/hole doping}

\begin{figure*}
\centering
\includegraphics[width=\textwidth]{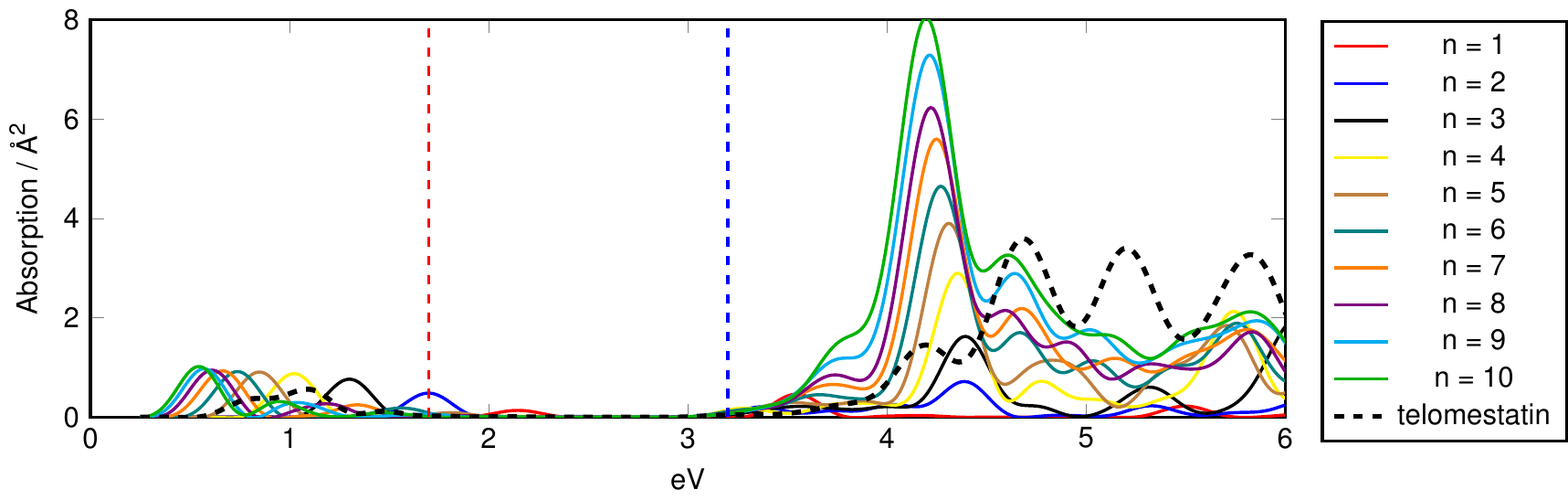}

\caption{Isotropic TD-ADSIC absorption spectra of hole-doped (mono-cationic) oligomeric oxazole chains. Vertical dashed lines indicate visible region. Spectra shifted by +0.25 eV.\cite{Doi2006}}
\label{fgr:abs_hdoped}
\end{figure*}

\begin{figure*}
\centering
\includegraphics[width=\textwidth]{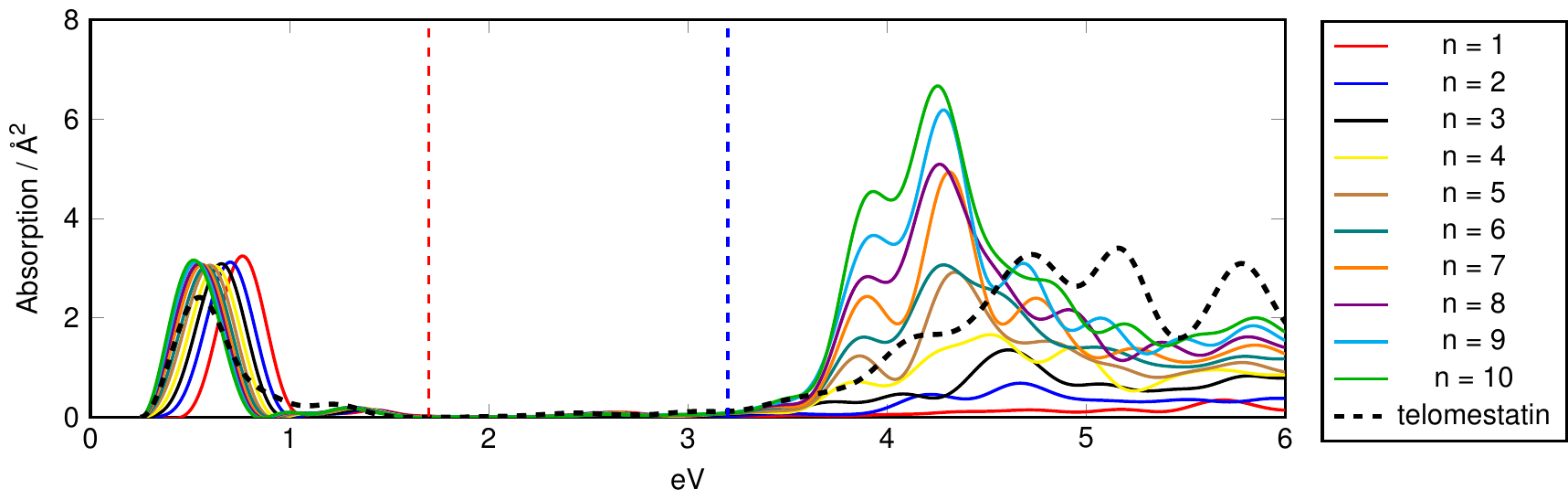}

\caption{Isotropic TD-ADSIC absorption spectra of electron-doped (mono-anionic) oligomeric oxazole chains. Vertical dashed lines indicate visible region. Spectra shifted by +0.25 eV.\cite{Doi2006}}
\label{fgr:abs_edoped}
\end{figure*}

In the cationic hole-doped species (Figure \ref{fgr:abs_hdoped}), the plasmon peak at $\sim$ 4.1 eV becomes slightly more intense but overall, this peak (and its satellites) maintain the shapes seen for the neutral species. In contrast, for the anionic electron-doped species (Figure \ref{fgr:abs_edoped}) the relative intensities of the peaks change with the low-energy satellite at $\sim$ 3.9 eV becoming more pronounced. This peak displays the same linear growth associated with the main peak in the neutral and hole-doped species but the growth of the main peak with chain length becomes more erratic. The telomestatin spectrum in the UV remains essentially unchanged with the same four peaks as seen for the neutral species, with only small relative intensity changes suggesting that the all-\textit{trans} linear species are more sensitive to doping effects in this region of the spectrum.

The largest changes in the absorption spectra for the doped species are seen in the IR region with electron-doping and from the visible into the IR for the hole-doped chains. These doping-induced peaks correspond to the formation of spin polaron (SP) gap states which have been well characterised in conjugated organic polymers.\cite{Bredas1984,Bredas1985} In polyanilines, (electro)chemical oxidation or reduction of the conducting polymer leads to creation of electronic transitions involving such gap states which vary in energy from the near IR to near UV. \cite{Tan2000}. A characteristic doped-polyaniline absorption associated with these states occurs at $\sim$ 0.7 eV in good agreement with the predictions for the polyoxazole chains.\cite{Stafstrom1987, Zheng1997}

\begin{figure}[h!]
   \centering
   \includegraphics[width=0.45\textwidth]{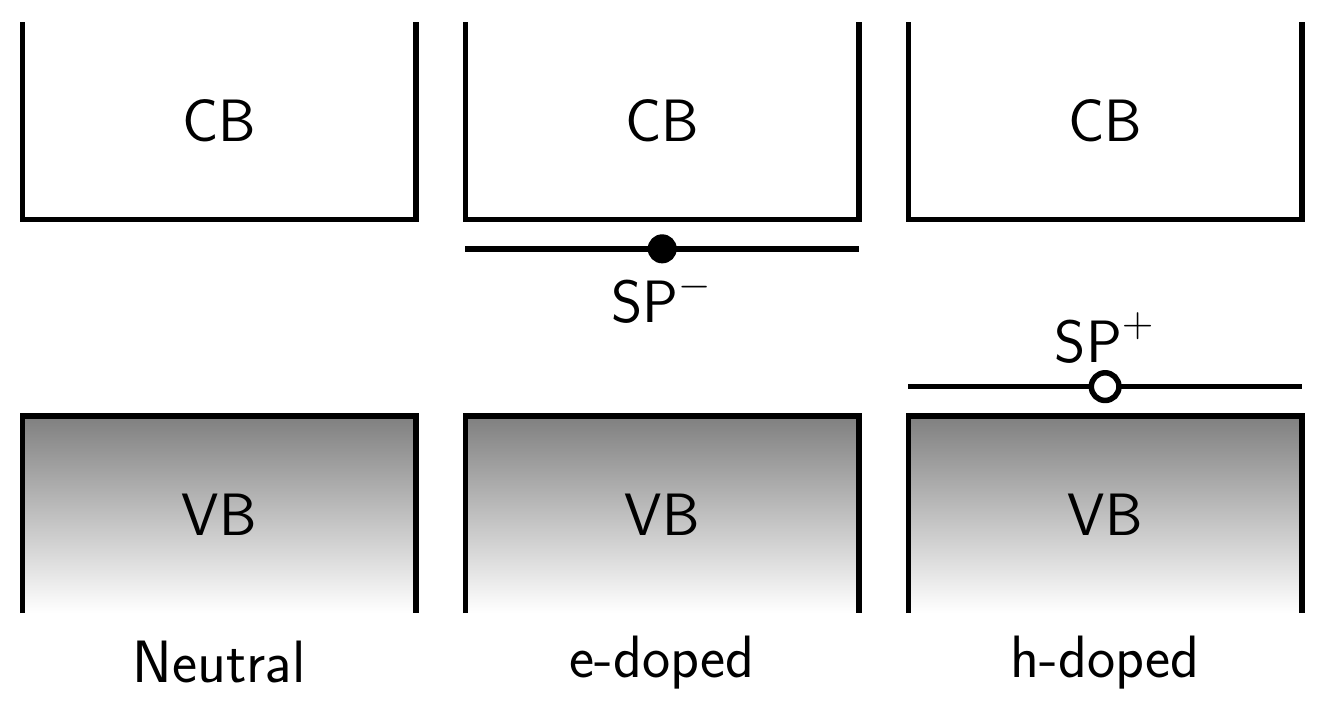}
   \caption{Schematic of semiconductor band gap between filled valence band (VB) and empty conduction band (CB) with spin polaron (SP) gap states caused by doping. Excess electron (hole) indicated by filled (empty) circle. Relaxation effects on gap not shown.}
   \label{fgr:gap_polarons}
\end{figure}

Figure \ref{fgr:gap_polarons} gives a schematic indication of the distribution of these states in the doped species relative to the band gap. In hole-doped species, excitation occurs from the top of the valence band (VB) to the unoccupied SP$^+$ state at bottom of gap. Similarly, in the electron-doped species, excitation occurs from the occupied  SP$^-$  the bottom of the conduction band (CB). 

In both cases, the excitation energy involved is substantially smaller than the band gap, hence the appearance of the peaks at much lower energies than seen in the spectra of the corresponding neutral species. VB $\rightarrow$ SP$^+$ and SP$^-$ $\rightarrow$ CB gaps are chain-length dependent and converge to less than 0.05 eV for n = 10 providing a rationale for the observed modulation of the associated absorption energies. 

Experimental measurement of polaronic optical excitation would necessarily involve nuclear relaxation through the electron/hole coupling to system phonons, which cannot be explicitly modelled by the current calculations due to the clamped nuclei (i.e. Born-Oppenheimer) approximation. However, it is felt that the current results reliably indicate the nature of the polaron states in electron-/hole-doped oligomers such as the polyoxazoles.

\section{Conclusions}

In the present work, a series of linear oligomers based on the polyoxazole structure of the natural product telomestatin were studied. Self-interaction corrected time-dependent density functional theory within the real-space/time-propagation formalism was used to investigate both the ground- and excited-state electronic structure of oligomers with lengths varying from one to ten oxazole rings.

It was found that the polyoxazoles display semiconductor character with a band gap converging to $\sim$2.5 eV as chain length increases with length-dependent formation of a discrete and compact conduction band (CB). On photo-excitation (or electron-doping) and population of the CB it is expected that they should display significant conductivity due to the large density of states in the low CB region. In addition, the oligomers display linear growth of the static polarisability which again marks them as potentially interesting as electrically conducting materials since this property can be systematically tuned by controlling oligomer length during synthesis.\cite{Gillet2021} Finally, we have shown that the neutral chains have polarisation-dependent maxima in their absorption spectra  at $\sim$4.1 eV, with length-dependent intensity suggestive of collective plasmonic excitations, and on electron/hole doping they develop additional absorption peaks in the IR region. 

The tunability of the oligomer's intrinsic electronic properties and combination of UV/IR absorption in the doped species suggests that the polyoxazoles could have important applications as optically transparent electronic materials in areas such as window-based photovoltaics .\cite{Sayem2022,Baron2022}

Substitutions of the oxygen atom in the individual heterocycles could also be envisioned as a route to fine-tuning the chains' properties. Heavier Group 16 elements such as Sulfur could be used in analogy with polythiophene to alter the band gap of the polyazole oligomers studied here. Beyond this, Selenium or even Tellurium can be expected to radically alter both the gap and the conduction band DOS as they are known to exert a significant effect on the structure of the lowest unoccupied molecule orbital (LUMO) in organic chromophores.\cite{Carrera2015,Kaloni2016} In addition, the greatly enhanced spin-orbit effects in the later Group 16 elements can be expected to produce large alterations in inter-system crossing probabilities allowing control of the optical emission properties of the oligomers.\cite{Carrera2015,Acharya2016}

Further studies of \textit{cis/trans} conformational effects on the DOS of the oligomeric chains along with alteration of the terminating groups or macrocyclisation would help to clarify the contributions coming from these sources. Similarly, modulation of the DOS and even suppression of the band gap to induce (semi-)metallic nature by $\pi$-stacking and supramolecular organisation in materials made from these oligomers can be expected and therefore investigation of these effects is warranted.\cite{Kaloni2017}

The convergence of the polaron gaps towards very low energies also suggests that the doped polyoxazoles should display significant electrical conductivity. As with the absorption spectra, it should be possible to tune the conductivity through modulating the chain length. Controlling the density of polaron-type charge carriers in organic materials has been demonstrated to have applications in fields as diverse as molecular sensing and photovoltaic materials.\cite{Tan2000,Tang2022} The stability/lifetime of the polaron states is expected to be reinforced by supramolecular effects such as $\pi$-stacking or hydrogen-bonding in bulk materials or thin films.\cite{Dumele2020} Further work is under way to investigate the structure and properties of such supramolecular assemblies involving the polyoxazoles and their group 16 analogues.

\section*{Acknowledgements}
The authors acknowledge financial support from the Industrial Biotechnology Innovation Centre (IBioIC, doctoral training grant BB/W059899/1). This work was also supported by national funds from the Portuguese FCT – Funda\c{c}\~{a}o para a Ci\^{e}ncia e a Tecnologia, I.P., within the projects UIDB/04564/2020 and UIDP/04564/2020. The authors thank the Laboratory for Advanced Computing (LCA) of the University of Coimbra, Portugal for technical support. The authors would also like to acknowledge the support of the Maxwell Compute Cluster funded by the University of Aberdeen. All 2D chemical structures in this work were created using Marvin version 23.1, ChemAxon (https://www.chemaxon.com).

\bibliography{telomestatin}
\bibliographystyle{unsrt} 

\end{document}